\documentclass[10pt,emptycopyrightspace]{ewsn-proc}
\usepackage{xspace}
\usepackage{balance}
\usepackage{graphicx}
\usepackage[hyphens]{url}
\usepackage{epstopdf}
\usepackage{microtype}
\usepackage{booktabs}
\usepackage[table,xcdraw]{xcolor}
\usepackage{caption}
\usepackage{subcaption}
\usepackage{amsmath}
\usepackage{enumitem}
\usepackage{amssymb}
\usepackage{pifont}
\iftrue 
	
\else
	
\fi

\newcommand{\PreserveBackslash}[1]{\let\temp=\\#1\let\\=\temp}
\newcolumntype{M}[1]{>{\PreserveBackslash\centering}m{#1}}
\newcolumntype{C}[1]{>{\PreserveBackslash\centering}p{#1}}
\newcolumntype{R}[1]{>{\PreserveBackslash\raggedleft}p{#1}}
\newcolumntype{L}[1]{>{\PreserveBackslash\raggedright}p{#1}}

\newcommand{\fakepar}[1]{\smallbreak\noindent}
\newcommand{\boldpar}[1]{\smallbreak\noindent\textbf{#1.}}

\newcommand{\ieee}{\mbox{IEEE~802.15.4}\xspace}

\newcommand{\btfive}{\mbox{BT\,5}\xspace}
\newcommand{\dcubee}{\mbox{D-Cube}\xspace}

\newcommand{\atomic}{\mbox{Atomic-SDN}\xspace}
\newcommand{\sixpp}{\mbox{6PP}\xspace}
\newcommand{\nrf}{\mbox{nRF52840}\xspace}

\usepackage{manyfoot}
\DeclareNewFootnote{A}
\DeclareNewFootnote{B}

\let\footnoteR\footnoteB
\let\footnote\footnoteA
\iftrue
    \newcommand{\mike}[1]{\footnoteR{{\color{red}\bf MB: #1}\color{red}}}
    \newcommand{\adnan}[1]{\footnoteR{{\color{red}\bf AA: #1}\color{red}}}
    \newcommand{\usman}[1]{\footnoteR{{\color{red}\bf UR: #1}\color{red}}}
    \newcommand{\alex}[1]{\footnoteR{{\color{red}\bf AS: #1}\color{red}}}
    \newcommand{\yichao}[1]{\footnoteR{{\color{red}\bf YJ: #1}\color{red}}}
    \newcommand{\george}[1]{\footnoteR{{\color{red}\bf GO: #1}\color{red}}}
    \newcommand{\carlo}[1]{\footnoteR{{\color{red}\bf CAB: #1}\color{red}}}
    \newcommand{\markus}[1]{\footnoteR{{\color{red}\bf MS: #1}\color{red}}}
\else
    \newcommand{\mike}[1]{}
    \newcommand{\adnan}[1]{}
    \newcommand{\usman}[1]{}
    \newcommand{\alex}[1]{}
    \newcommand{\yichao}[1]{}
    \newcommand{\george}[1]{}
    \newcommand{\carlo}[1]{}
    \newcommand{\markus}[1]{}
\fi

\begin{document}

% Title
\title{
% \vspace{-1.50mm}
6TiSCH++ with Bluetooth\,5 and Concurrent Transmissions
% \vspace{-5.00mm}
}

% Authors (blinded)
\iffalse
\author{
    \alignauthor Anonymous Author(s) \\
	 EWSN 2021 -- Submitted Paper \#32
}
\else
\author{
\alignauthorpage Michael Baddeley$^{*}$,~Adnan Aijaz$^{*}$,~Usman Raza$^{*}$,~Aleksandar Stanoev$^{*}$,~Yichao Jin$^{*}$,\\~Markus Schu{\ss}$^{\dagger}$,~Carlo Alberto Boano$^{\dagger}$~and~George Oikonomou$^{\ddagger}$\\
    \vspace{1mm}
    \affaddr{$^{*}$Toshiba Europe Ltd., $^{\dagger}$Graz University of Technology, $^{\ddagger}$University of Bristol}\\
    \email{\{m.baddeley; a.aijaz; u.raza; a.stanoev; y.jin\}@toshiba-bril.com}\\
    \email{\{markus.schuss; cboano\}@tugraz.at, g.oikonomou@bristol.ac.uk}
}
\fi

\maketitle

\begin{abstract}
Targeting dependable communications for industrial \mbox{Internet} of Things applications, IETF 6TiSCH provides mechanisms for efficient scheduling, routing, and forwarding of IPv6 traffic across low-power mesh networks. Yet, despite an overwhelming body of literature covering both \mbox{centralized} and distributed scheduling schemes for 6TiSCH, an effective control solution for large-scale multi-hop mesh networks remains an open challenge. Our paper fills this gap with a novel approach that eliminates much of the routing and link-layer overhead incurred by centralized schedulers, and provides a robust mechanism for data dissemination synchronization within 6TiSCH. Specifically, we leverage the physical layer (PHY) switching capabilities of modern low-power wireless platforms to build on recent work demonstrating the viability of Concurrent Transmission (CT)-based flooding protocols across the Bluetooth\,5 (\btfive) PHYs. By switching the PHY and MAC layer at runtime, we inject a \btfive-based CT flood within a standard \ieee TSCH slotframe, thereby providing a reliable, low-latency scheme for 6TiSCH control messaging. We present an analytical model and experimental evaluation showing how our solution not only exploits the \btfive high data-rate PHYs for rapid data dissemination, but can also provide reliable 6TiSCH association and synchronization even under external radio interference. We further discuss how the proposed technique can be used to address other open challenges within the standard.

% \george{Overall very well-written abstract - worth checking for any possible word count-related limitations - it feels a bit long. Some tech detail could perhaps be omitted}
\end{abstract}

\iffalse
\fi

\section{Introduction}
\label{sec:introduction}

The Time Slotted Channel Hopping (TSCH) MAC layer option in the \ieee-2015 standard
% ~\cite{ieee_802154} 
provides an alternative to asynchronous channel access, which can suffer from contention issues that limit service guarantees. TSCH introduces scheduled communications across time and frequency: thus providing a mechanism to escape external interference, follow optimal transmission schedules, and improve network scalability and throughput by allowing co-located nodes to transmit on orthogonal channels. However, TSCH doesn't specify how a schedule should be built and maintained, instead leaving it to higher layers to define either a static schedule, or allow nodes to contest timeslots. IETF 6TiSCH~\cite{ietf_6tisch} was created to bridge this gap. Besides defining default functions for creating and disseminating a schedule, 6TiSCH establishes mechanisms for deterministic IPv6 routes across the mesh. In this fashion, 6TiSCH is able to offer service guarantees through efficient allocation of radio and network resources. Yet, to-date, there exist relatively few examples of successful real-world deployments. Although large-scale mesh networks have found application in smart metering~\cite{tepco_landys}, within industrial use-cases mesh is often merely employed as range extender for cellular systems~\cite{dcc_comshub}. 
% MB: Agree. Have simply taken it out as the sentence after says pretty much the same thing anyway.
% Simply put, for many use-cases and customers, low-power wireless mesh is a sub-optimal solution that introduces excessive complexity and limits scalability\george{This sentences comes across somewhat dogmatic and strong - suggest reviewing}. 
While the efforts of 6TiSCH have been considerable, it is impossible to ignore the fact that maintaining mesh networks is still incredibly complex depending on the environment and application, while other wireless solutions tend to `just work'.

Subsequently, there has been considerable interest in developing communication protocols based on Concurrent~Transmissions~(CT), where nodes synchronously transmit in-contention with their neighbors. Although conventional wisdom would suggest that contending transmissions will collide and will not be demodulated at the receiver, a number of physical layer (PHY) effects~\cite{leentvaar76capture,liao2016revisiting} and considerable MAC layer redundancy ensure high probability of a correct reception. Multiple editions of the EWSN dependability competition~\cite{boano17competition} have also shown that CT-based flooding protocols outperform conventional approaches across a number of key performance metrics, even under high levels of external radio interference. Crucially, as they rapidly flood the network with high probability, CT protocols eliminate reliance on distributed control signaling and therefore address the routing and scheduling complexity seen in standard mesh solutions. 

It is this simplicity that makes CT protocols a promising candidate for supporting 6TiSCH networks. To-date, the weight of 6TiSCH literature has focused on \emph{distributed} as opposed to \emph{centralized} scheduling mechanisms, which face considerable research challenges, such as significant jitter over multi-hop links and topological funneling effects near a central coordinator. Indeed, the lack of means to efficiently disseminate centrally-computed control signaling poses a significant obstacle to the adoption of concepts such as Software Defined Networking (SDN) -- currently defined within the standard~\cite{ietf_6tisch}.

\begin{figure}[!h]
	\centering
% 	\vspace{+1.00mm}
	\includegraphics[width=.68\columnwidth]{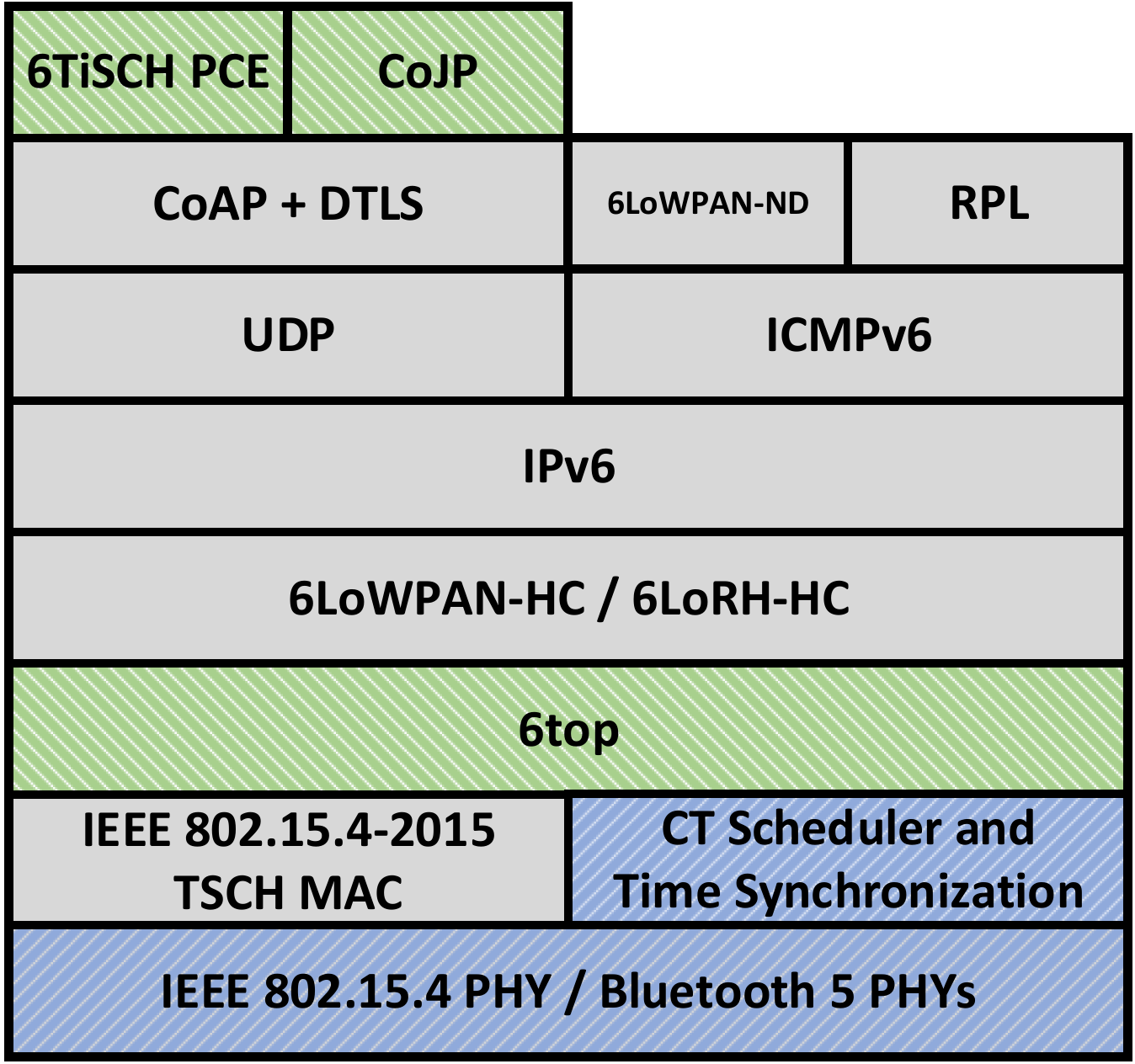}
	\vspace{-2.00mm}
	\caption{\sixpp stack with respect to the 6TiSCH standard.}
% Synchronization and centralized control messaging are handled via a \btfive-capable CT scheduler.}
	%\vspace{-0.1m}
	\label{fig:6pp_stack}
\end{figure}

\boldpar{Our contributions} We propose 6TiSCH++ (\sixpp), in which we exploit recent advances in CT protocols~\cite{alnahas2019concurrentBLE5} alongside multi-PHY capabilities of modern low-power radios~\cite{nrf52840_productsheet}. By utilizing the \btfive PHYs (though we recognize that \ieee-based CT is also a valid option) \sixpp provides a suite of high-data rate and coded PHY options that benefit from recent understanding of the physical layer impact on CT protocols~\cite{baddeley2020impact}. Crucially, our solution disseminates network configuration and synchronization information over the CT layer, and thus eliminates much of the routing and link-layer signaling overhead that hinders current 6TiSCH solutions. Fig.~\ref{fig:6pp_stack} demonstrates how \sixpp fits neatly within the standard 6TiSCH stack. We argue that CT-flooding over the high data-rate (1M and 2M) \btfive PHYs addresses the challenge of how to provide rapid, reliable distribution of 6TiSCH control messaging, while switching to the \btfive coded PHYs can provide robust network synchronization even under external interference.

\boldpar{Outline of this paper} After providing a primer on relevant aspects of 6TiSCH and CT, Sect.~\ref{sec:background} discusses the impact of \sixpp with respect to related works. The rest of this paper is structured as follows.

\begin{itemize}
    \item Technical aspects \sixpp are outlined in Sect.~\ref{sec:approach}.
    \item In Sect.~\ref{sec:experimentation} we validate our solution on the \dcubee public testbed~\cite{schuss17competition} and demonstrate how \sixpp is capable of disseminating both 6TiSCH and RPL control signaling across the entire mesh with minimal latency.
    % MB: Agree, so have renamed this subsection "Outline"
    % \george{Certainly valuable and important point to mention, but I would not strictly call it a contribution}.
    \item Sect.~\ref{sec:simulation} provides simulation-based evaluation demonstrating how encapsulating a CT-based flood within the slotframe can significantly improve data dissemination latency and reliability within a 6TiSCH network.
\end{itemize}

\smallbreak\noindent Finally, Sect.~\ref{sec:conclusions} gives directions and insights on how this work may be taken forward in future research as well as standardization activities, and concludes this paper. 

\begin{figure}[ht]
	\centering
	\includegraphics[width=.9\columnwidth]{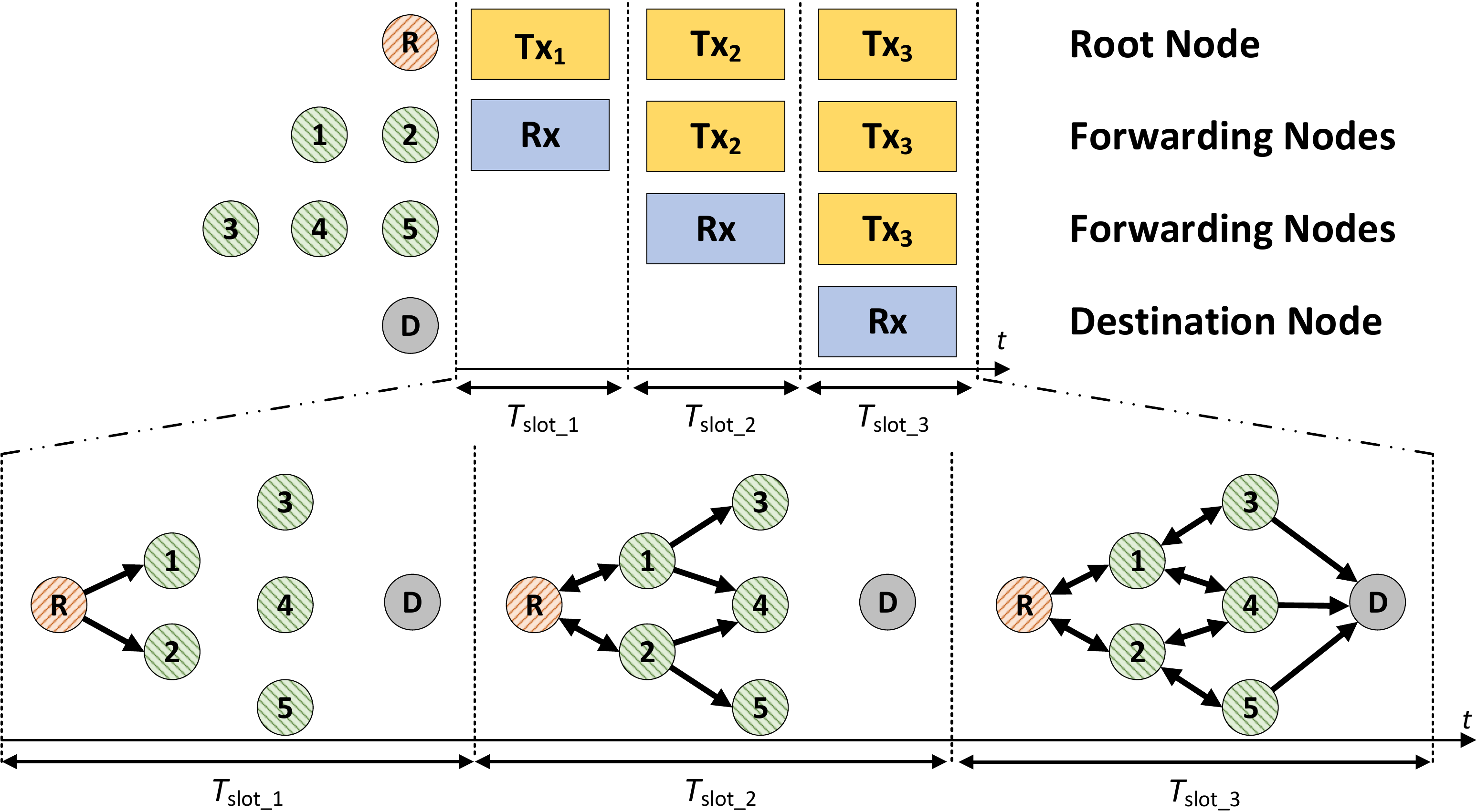}
	\caption{Time-triggered CT flood.~\cite{lim2017competition}.}
	\vspace{-0.2cm}
	\label{fig:ct_example}
\end{figure}

% \todo{Enumerate the practical uses of this technique. It could be used to support X [x] in blah, Y [y] in blah, and Z [z] in blah.}

% \george{Overall, I feel the introduction is very well-written and puts all key points across nicely.

% I think it is too long though, I tend to force my introductions to end in page 1, column 2.

% The ``Our contributions'' section needs revised IMHO. I feel it mixes contribution, with details about how the work is presented and the paper structure. I think the contribution statement is from ``We propose...'' to ``synchronization mechanism.'' \textcolor{blue}{MB: Cheers, George, this helps immensely. Will do this tomorrow.}}

\begin{figure*}[!t]
	\centering
	\includegraphics[width=.9\textwidth]{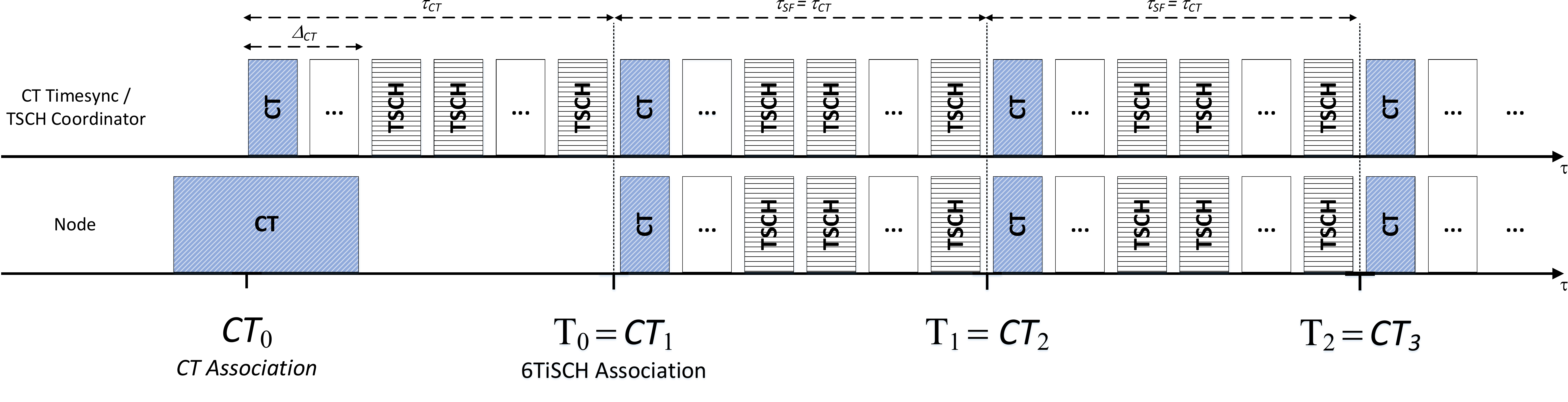}
	\vspace{-3.00mm}
	\caption{In 6TiSCH++ IEEE 802.15.4 EBs are flooded over \btfive-based CT to quickly synchronize and associate 6TiSCH.}
	\vspace{+4.00mm}
	\label{fig:our_approach}
\end{figure*}
\begin{figure}[!b]
    % \vspace{-0.3cm}
	\centering
	\includegraphics[width=1\columnwidth]{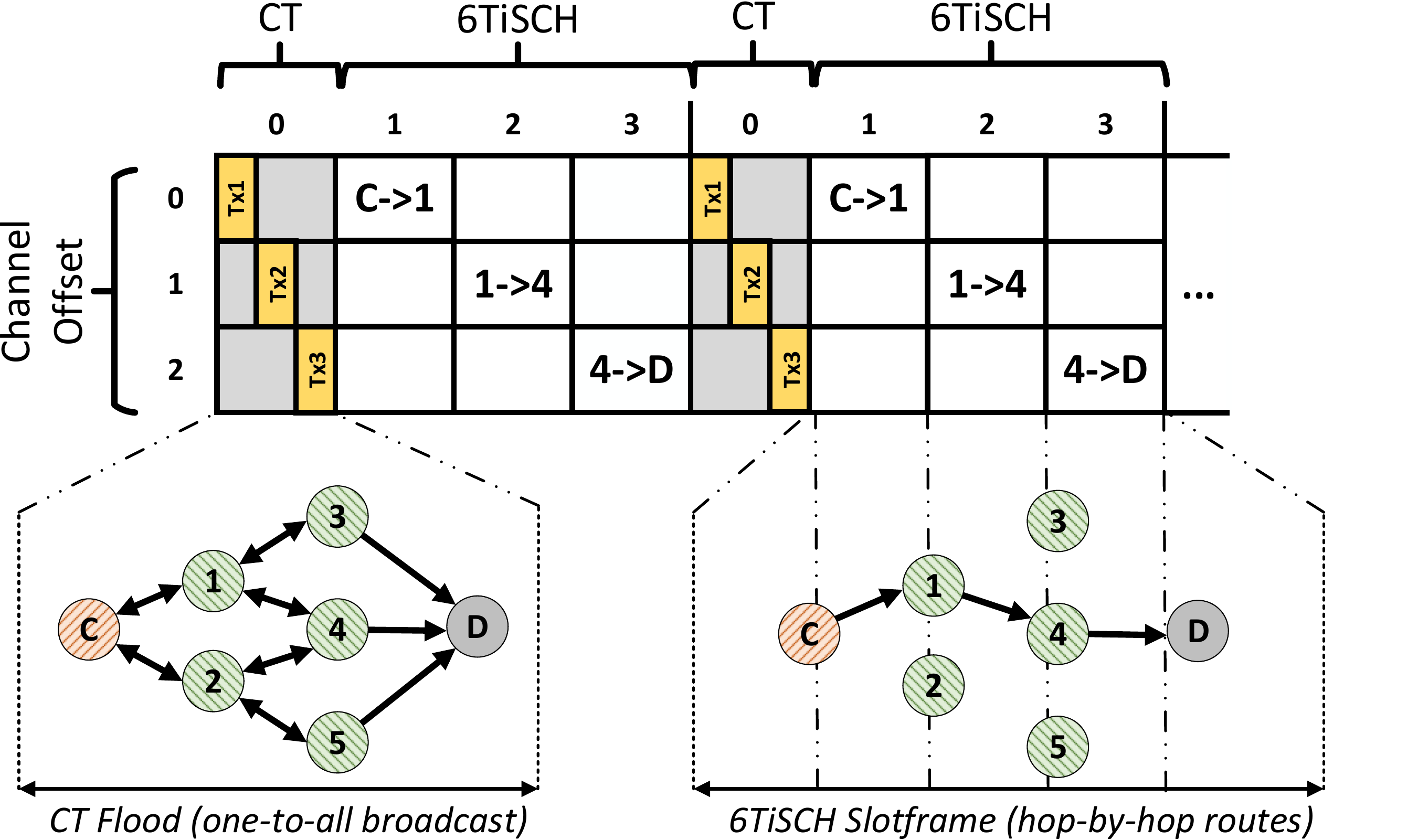}
	\caption{In \sixpp CT floods are interleaved within the 6TiSCH slotframe.}
% 	\vspace{-0.6cm}
	\label{fig:6pp_example}
\end{figure}

\section{Background and Related Work}
\label{sec:background}
We provide a brief overview of key aspects in both 6TiSCH and CT that are relevant to this paper. For a detailed examination of 6TiSCH we direct the reader toward~\cite{ietf_6tisch,vilajosana2019ietf}, whereas for a survey of CT-based protocols and the underlying physical layer phenomena underpinning CT we refer to~\cite{alnahas2019concurrentBLE5,baddeley2020impact,zimmerling20synchronous, jakob20experimental}.%schaper2019truth

%-------------------------------------------------------------------------------%
\boldpar{Centralized mesh control} 6TiSCH provides mechanisms for both \emph{centralized} and \emph{distributed} scheduling of the TSCH slotframe resources. Specifically, 6TiSCH differentiates between  centrally allocated \emph{hard} cells, and \emph{soft} cells that are negotiated between neighboring nodes on a hop-by-hop basis across the underlying layer-3 topology. 
% Yet while centralized approaches offer efficient control and global network overviews (indeed, this is the fundamental draw of the Software Defined Networking (SDN)~\cite{ietf_sdn} concepts that are part of the 6TiSCH standard), 
Yet, there are acute complexities in managing the signaling required to centrally provision networking and radio resources across a multi-hop mesh network. In addition to unreliable multi-hop links, the tree-like graphs formed by the commonly employed RPL routing protocol~\cite{ietf_rpl_2012} result in funneling effects that can cause severe delays and jitter near the root node. Indeed, \emph{downward} messaging -- i.e., multi-hop messages \emph{from} the root node \emph{to} nodes further down the tree -- are an historic weakness in mesh networks~\cite{iova2016rpl}. 

% While there are many works that have explored innovative 6TiSCH scheduling algorithms~\myciteme, a truly efficient and scalable means of centralized control has (until now) remained out of reach. Much of the 6TiSCH literature has therefore focused on \emph{distributed} means of allocating TSCH slotframe resources~\myciteme. 

%-------------------------------------------------------------------------------%
\boldpar{Synchronization, bootstrapping, and routing} Within TSCH, synchronization is achieved through the inclusion of Information Elements (IEs) within Enhanced Beacons (EBs) and Keep Alive (KA) messages. Specifically, EBs are periodically sent by neighbors to synchronize joining nodes through the IE, as well as provide information on the configuration necessary to bootstrap and securely join the network. Once joined, IEs are again employed within KAs to maintain synchronization and compensate drift~\cite{vilajosana2019ietf}. 

While RPL supports lightweight mechanisms for joining a tree-like graph for sending messages \emph{upwards} towards the root, \emph{downward} messaging involves a two-way handshake across multiple hops, with nodes declaring their existence through Destination Advertisement Object (DAO) messages, which are repeatedly sent to the root until the node receives an acknowledgment (DAO-ACK). As a network scales, DAOs sent by nodes further away are therefore subject to greater uncertainty, while poor links at bridging nodes can potentially occlude whole subsections of the network.

Crucially, the periodic nature of EBs and KAs means they represent a considerable portion of the overall 6TiSCH control messaging overhead, while the unreliable nature of RPL downward messaging can result in multiple DAO retransmissions to the root node in the event of missed DAO-ACKs -- a significant issue in networks with unreliable links, or interference scenarios~\cite{iova2016rpl}.

%-------------------------------------------------------------------------------%
\boldpar{Concurrent Transmissions (CT)} By \emph{intentionally} scheduling nodes to transmit in-contention with their neighbors, CT-based protocols rapidly disseminate packets across a multi-hop mesh. Fig.~\ref{fig:ct_example} demonstrates how time-triggered \emph{back-to-back} node transmissions~\cite{lim2017competition} are initiated by the arrival of a correct reception from neighboring nodes. Packets are thus disseminated within theoretical minimal bounds on latency (as dictated by the data rate of the underlying PHY). 
However, while the majority of CT literature has been based on the 2.4\,GHz \ieee OQPSK-DSSS PHY~\cite{zimmerling20synchronous}, 
% CT has been shown to be applicable over sub-GHz \ieee~\cite{liao16toward}, UWB~\cite{lobba20concurrent}, and LoRa~\cite{liao17lora}. Crucially, 
recent research has demonstrated CT over the \btfive PHYs~\cite{alnahas2019concurrentBLE5,jakob20experimental}. With modern low-power wireless platforms supporting multiple PHY options on a single chip (such as the Nordic~nRF52840~\cite{nrf52840_productsheet}, which supports both the \ieee OQPSK-DSSS and all four \btfive PHYs), there is emerging interest in the development of CT protocols which leverage these multi-PHY features~\cite{baddeley2020impact}. Furthermore, as these chipsets allow physical layer switching with \emph{no additional radio overhead}, CT-based scheduling architectures~\cite{ferrari2012low,jacob2019baloo,baddeley2019atomic} (which provide time-triggered scheduling services for a variety of CT-based protocols) are uniquely placed to offer multi-PHY scheduling solutions as they abstract the complexities of writing bare-metal CT-based protocols.

\boldpar{Related literature} While previous literature has proposed integration with CT-based flooding protocols as a means for reliable and low-latency data dissemination within \ieee~TSCH networks~\cite{baddeley2019atomic,chang2018constructive} there has been limited practical demonstration. Gomes et al.~\cite{gomes2016reliability} have proposed \ieee CT-based flooding within a TSCH slotframe as part of a solution submitted to the EWSN Dependability Competition~\cite{boano17competition}. However, this work did not target a 6TiSCH implementation (which would not have been possible on the employed legacy hardware~\cite{polastre05telos} used in initial editions of the competition). More recently, and particularly relevant to this paper, Istomin et al.~\cite{istomin2019route} have examined the use of CT to carry \emph{downward} RPL traffic in asynchronous CSMA/CA \ieee networks, while retaining routing-based transmission for \emph{upward} traffic; thus successfully building on the strengths of both approaches. Furthermore, the authors of~\cite{aijaz2019gallop} demonstrate the applicability of \btfive CT as a mechanism for single-hop cooperative transmissions and multi-hop time-synchronization alongside a secondary optimized transmission schedule, drawing on similar reasoning to the arguments presented in this paper. However, this approach is neither tailored to 6TSCH nor addresses the considerable challenges of 6TSCH integration. Finally, a recent survey~\cite{zimmerling20synchronous} speculates on current research challenges, decoupling CT-protocols using dual-processor platforms, and future routes to standardization, thereby supporting the fundamental reasoning behind this work.  

To the best of our knowledge there has been no study proposing the use of \btfive CT-based data dissemination in 6TiSCH networks, and ours is the first work to successfully demonstrate encapsulation of a CT flood within the \ieee TSCH slotframe.

\begin{figure*}[t]
    \centering
    \begin{subfigure}[t]{1\columnwidth}
    	\centering
    	\includegraphics[width=0.75\columnwidth]{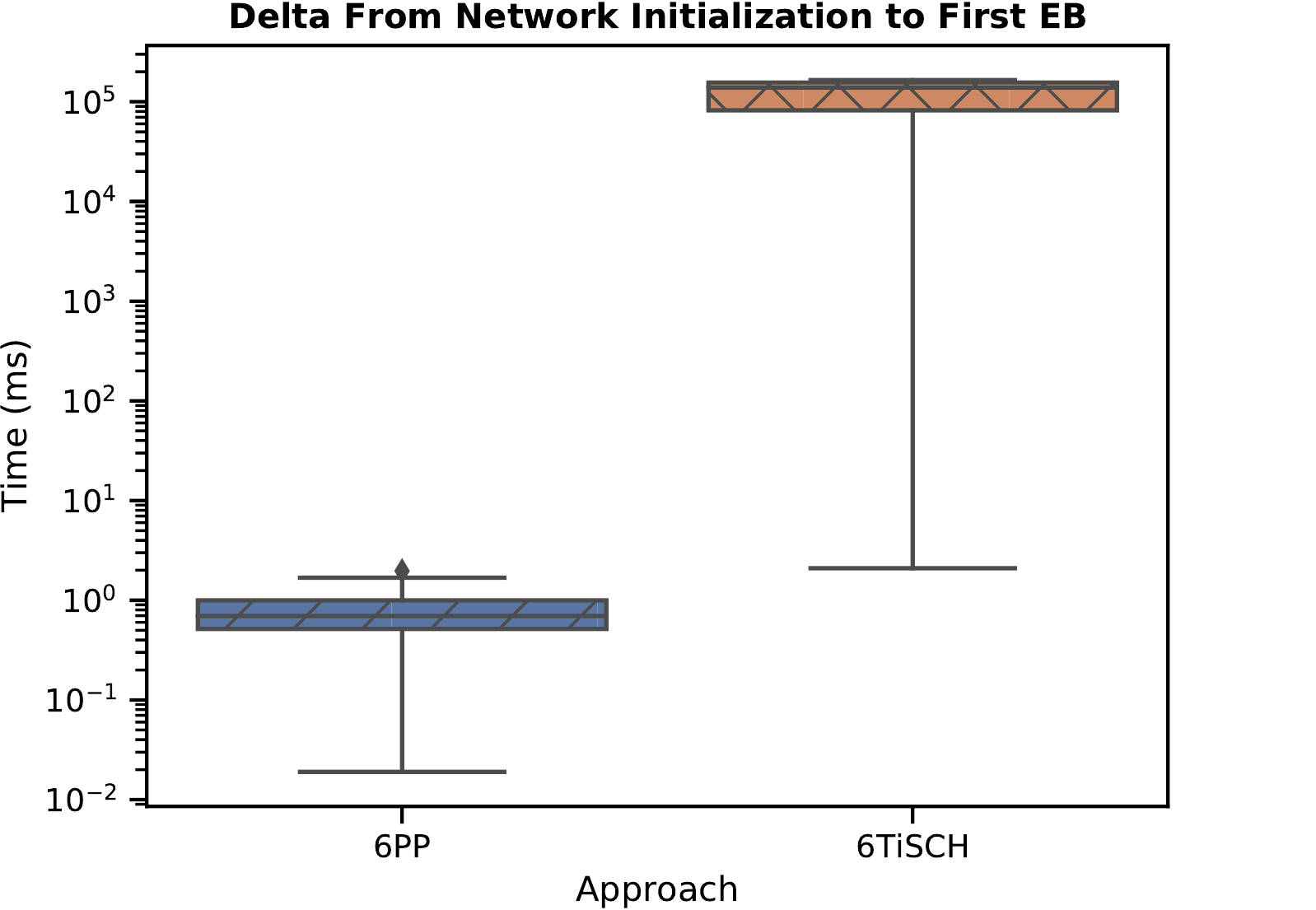}
    	\caption{Time taken for all D-Cube nodes to associate to the first EB.}
    	\label{fig:eb_delta}
    	\vspace{0.2cm}
    \end{subfigure}
    \begin{subfigure}[t]{1\columnwidth}
    	\centering
    	\includegraphics[width=0.75\columnwidth]{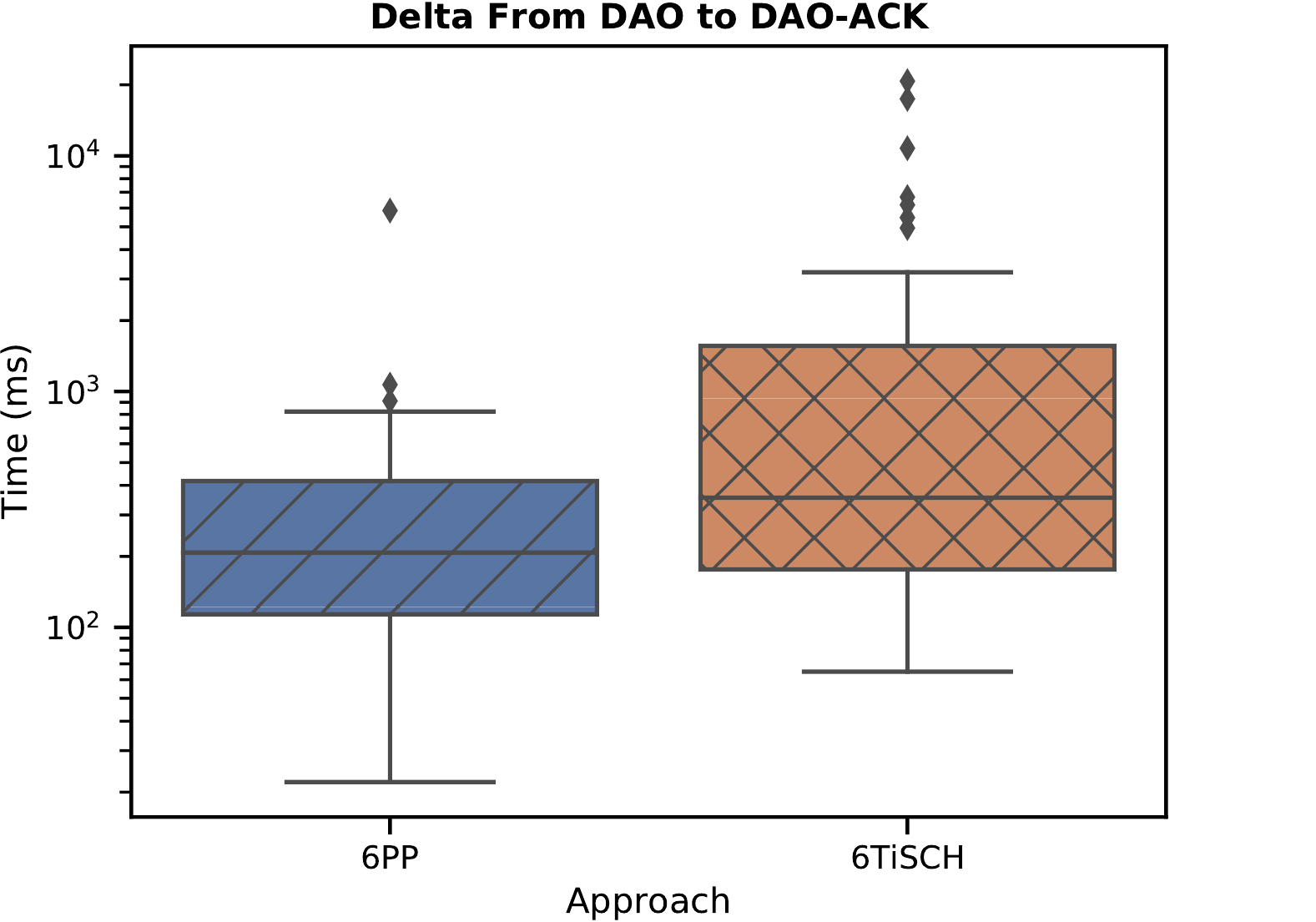}
    	\caption{Timedelta from sending a DAO to receiving a DAO-ACK.}
    	\label{fig:dao_delta}
    \end{subfigure}
    \vspace{-0.2cm}
    \caption{Time taken to (a) associate and (b) establish \emph{downwards} routes on D-Cube (48 nodes over $\approx$10-hops).}
    \vspace{-0.5cm}
    \label{fig:eb_dao_deltas}
\end{figure*}

\section{6TiSCH++: Concurrent MAC Scheduling}
\label{sec:approach}
\sixpp is designed around the careful and cooperative interleaving of periodic CT floods within a 6TiSCH slotframe -- demonstrated in Figs.~\ref{fig:our_approach} and~\ref{fig:6pp_example}. This dual-MAC approach schedules CT floods within dedicated `free' slots of a slotframe, while remaining slots are delegated to the 6TiSCH scheduler. By encapsulating a CT flood within the 6TiSCH slotframe we switch both the MAC and PHY layer in \emph{real-time}, with only a 40ms radio ramp-up time as overhead~\cite{nrf52840_productsheet}.

Fig~\ref{fig:6pp_example} shows a high-level example of how \sixpp is capable of scheduling time-triggered CT floods (as per Fig.~\ref{fig:ct_example}) alongside existing 6TiSCH layer-2 links. While this figure demonstrates three back-to-back transmissions in the CT slot, \textit{by employing the \btfive PHYs} (specifically, the high data rate uncoded options) \textit{it is possible to have many transmissions in the time it takes to complete a single 6TiSCH 10ms slot}. This key benefit of \sixpp is further explored in Sect.~\ref{sec:simulation}.

\sixpp employs a variation on the standard 6TiSCH minimal configuration~\cite{ietf_6min_rfc8180}. 6TiSCH minimal bootstraps the network with a basic schedule that provides a single shared slot for all data, synchronization, and advertisement. This allows nodes to associate to the network, maintain synchronization, as well as transmit and receive data. In this manner, a slotframe size of 1 (i.e., back-to-back TSCH slots) would be equivalent to slotted CSMA, while increasing the slotframe size retains the single shared slot but leaves the remainder of the slotframe empty -- saving energy. \sixpp denotes a new CT designated link type at the start of the slotframe and reserves the number of equivalent 10\,ms TSCH slots spanned by $\Delta_{CT}$. However, rather than leaving the remainder of the slotframe free (as in 6TiSCH minimal), \sixpp populates the rest of the slotframe with $N$ shared slots. We stress however, that although 6TiSCH minimal is used to validate the fundamental approach of \sixpp (in Sect.~\ref{sec:experimentation} we compare against a 6TiSCH minimal configuration with slotframe size 1), the \sixpp CT slot reservation mechanism could be applied to any other (dynamic) 6TiSCH scheduler. In such a case, hard cells would be allocated upon joining the network to ensure that the \sixpp slots are kept free for scheduled CT floods, and we propose this as a future research topic in Sect.~\ref{sec:conclusions}.

Fig.~\ref{fig:our_approach} shows the association and bootstrapping process of \sixpp. A single node is designated as \emph{both} the CT timesync and the 6TiSCH coordinator. The joining and synchronization information contained within EB messages, usually sent on a hop by hop basis, is instead disseminated rapidly across the entire mesh through the CT flood. Once synchronized with the CT timesync, joining nodes set the reference time of the \ieee TSCH timer from the reference time captured by the CT scheduler ($CT_{0}$), and set a CT flooding period equivalent to the duration of the pre-configured 6TiSCH slotframe duration such that $\tau_{CT} = \tau{SF}$. Successful CT association subsequently starts the TSCH association process, while disassociation from the CT network likewise disassociates the node from the TSCH network (and RPL DAG). Drift correction is performed every flooding period by compensating the TSCH reference time from the CT scheduler. In this manner, \sixpp removes the need for KA messages, and further reduces the overall control signaling overhead within the mesh.

Finally, building on the similar approach taken in~\cite{istomin2019route}, \sixpp disseminates RPL DAO-ACK messages over CT floods. This allows \sixpp to reliably and rapidly establish the \emph{downward} routes, which is a well-known challenge in wireless mesh networks~\cite{iova2016rpl}.

% \todo{Concurrent scheduling of CT and 6TiSCH.}

% \todo{CT dissemination of control signaling.}

% \todo{CT-triggered TSCH association and slotframe alignment.}

% \todo{TSCH timer compensation.}

\begin{table}[t]
    \vspace{0.2cm}
	\renewcommand{\arraystretch}{1.0}
	\centering
	\footnotesize
    \begin{tabular}{ l c c }
    \toprule
      	\bfseries Approach & \bfseries Reliability (\%) & \bfseries Mean Latency (ms)\\ \midrule
        6PP (No Interf.) & 100 & 250.60 \\
        6TiSCH (No Interf.) & 100 & 403.96 \\
        6PP (With Interf.) & 99.54 & 329.58 \\
        6TiSCH (With Interf.) & 98.63 & 527.64 \\
    \bottomrule
    \end{tabular}
    \caption{Reliability and latency for 6PP and 6TiSCH in \dcubee's 20-node dense data dissemination scenario (\emph{with} and \emph{without} external narrowband interference).}
    \label{table:dcube_1hop}
\end{table}
\begin{figure}[t]
	\centering
	\includegraphics[width=0.7\columnwidth]{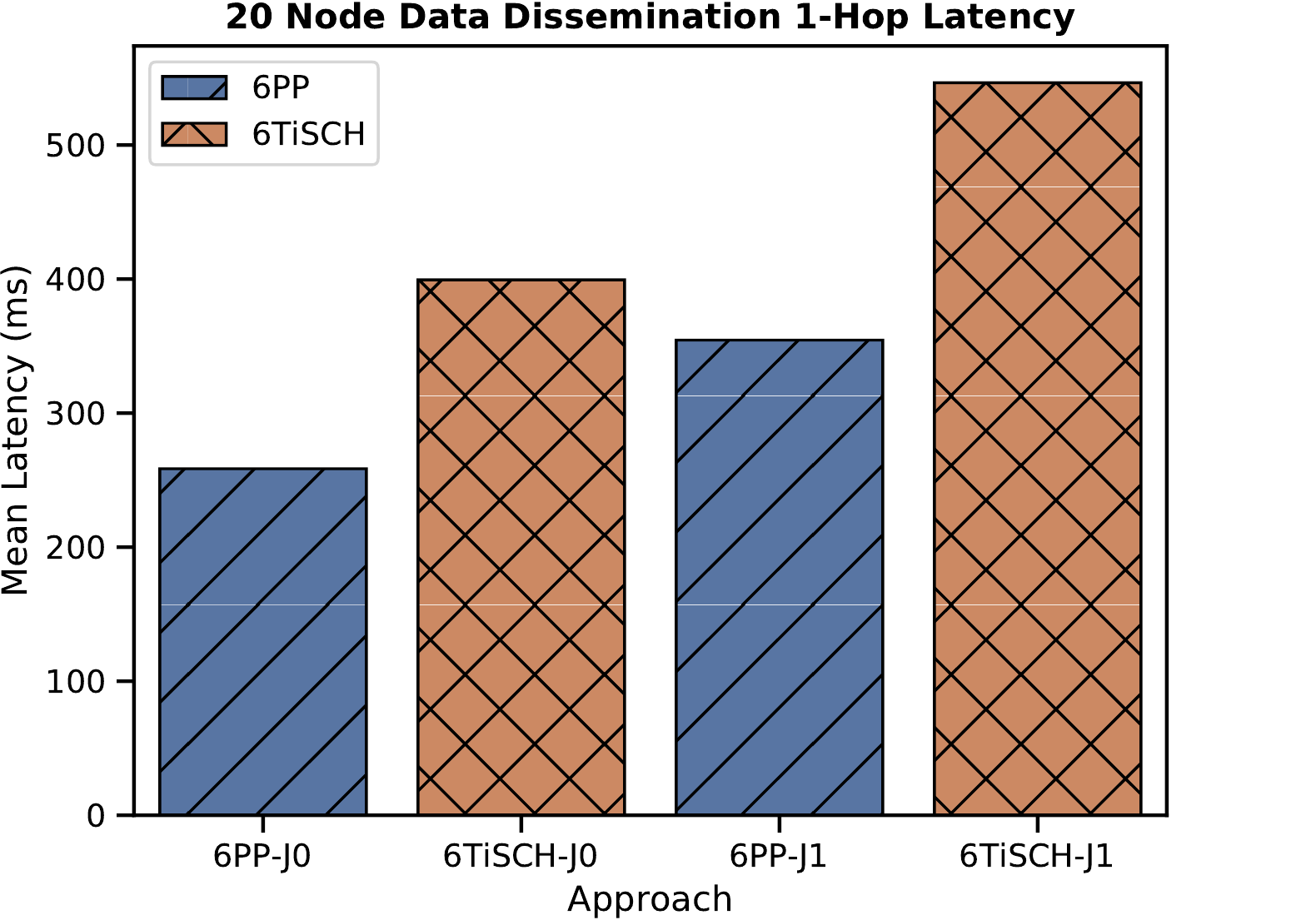}
	\caption{Mean latency from Table~\ref{table:dcube_1hop} -- \emph{without} (J0) and \emph{with} (J1) external narrowband interference.}
	\label{fig:dcube_1hop}
	\vspace{-0.2cm}
\end{figure}
\begin{figure*}[t]
    \centering
    \begin{subfigure}[t]{0.32\textwidth}
    	\centering
    	\includegraphics[scale=0.2]{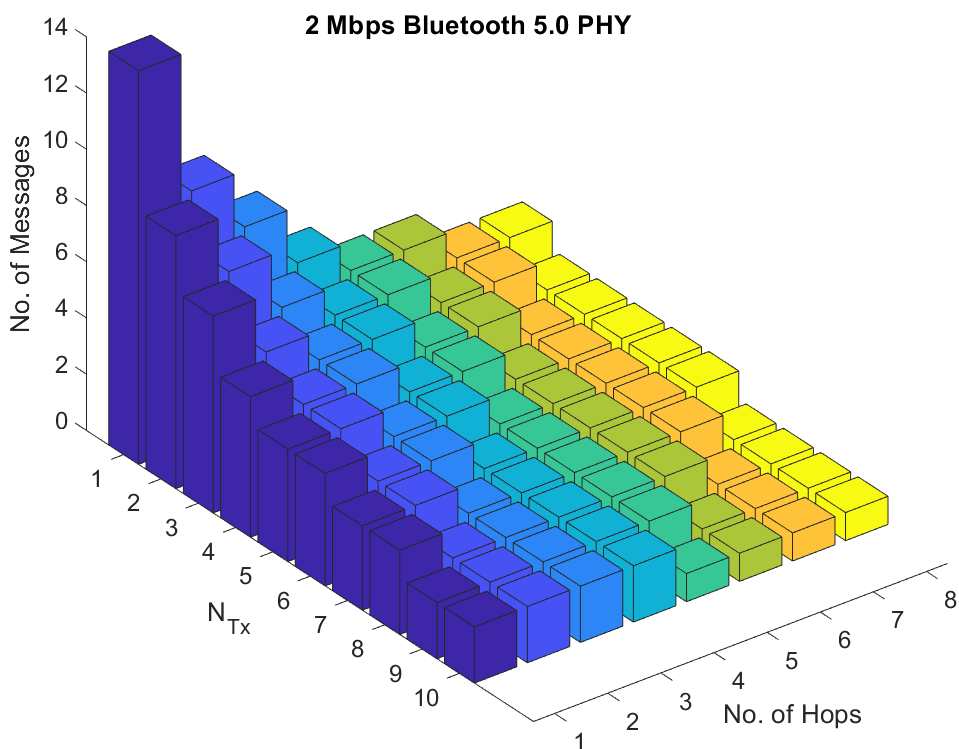}
    	\caption{}
    	\vspace{0.2cm}
    	\label{fig:sim_latency_2M}
    \end{subfigure}
    \begin{subfigure}[t]{0.32\textwidth}
    	\centering
    	\includegraphics[scale=0.2]{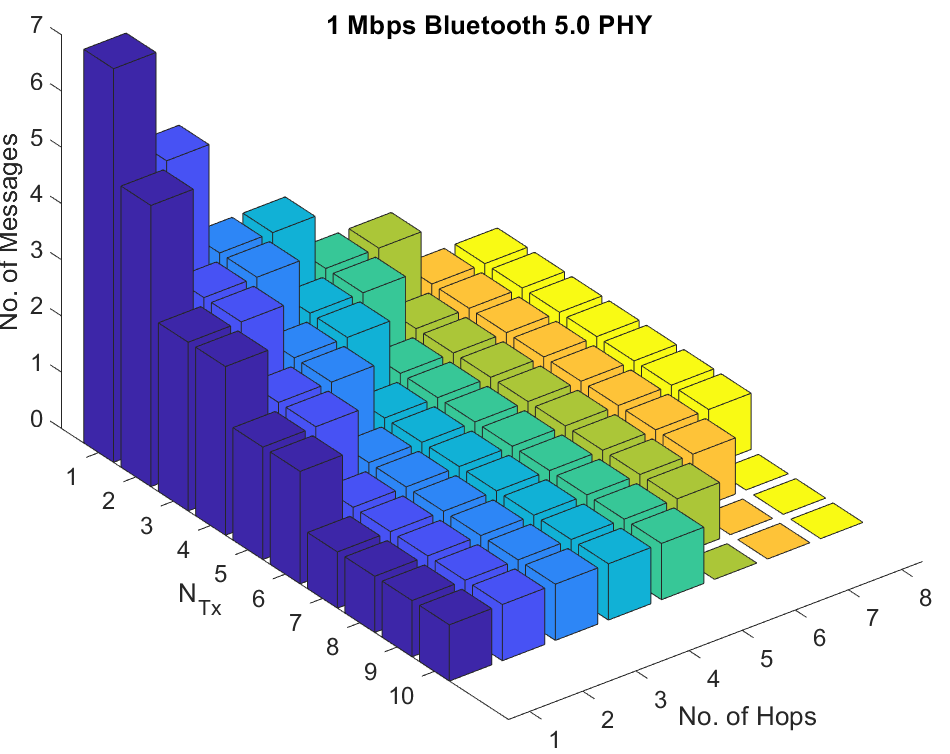}
    	\caption{}
    	\vspace{0.2cm}
    	\label{fig:sim_latency_1M}
    \end{subfigure}
    \begin{subfigure}[t]{0.32\textwidth}
    	\centering
    	\includegraphics[scale=0.2]{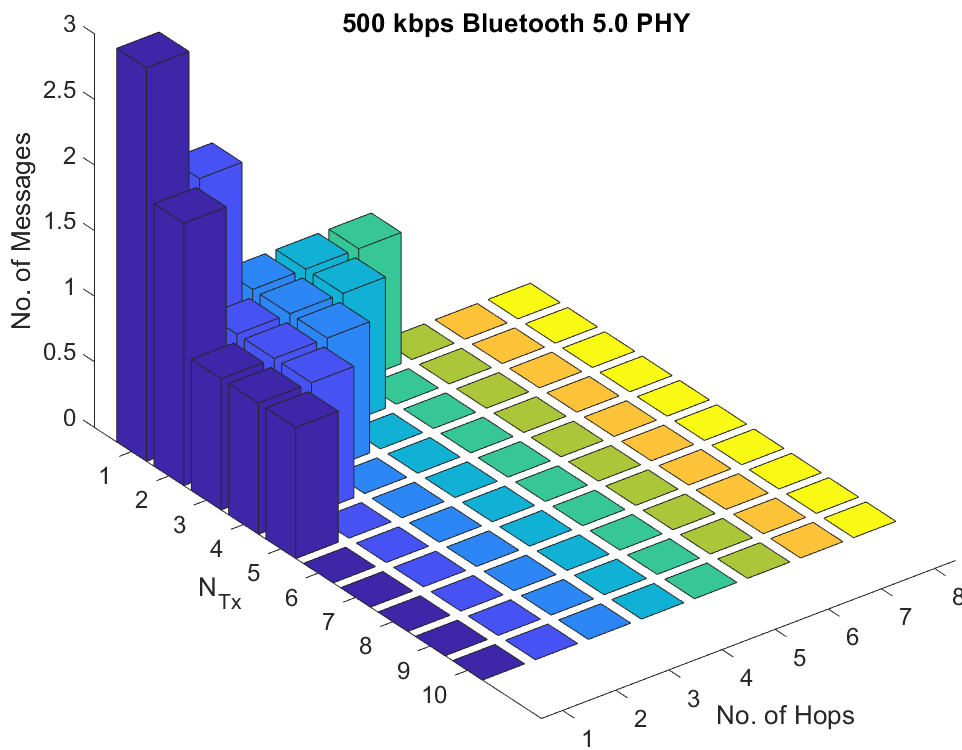}
    	\caption{}
    	\vspace{0.2cm}
    	\label{fig:sim_latency_500K}
    \end{subfigure} 
    \begin{subfigure}[t]{0.32\textwidth}
    	\centering
    	\includegraphics[scale=0.2]{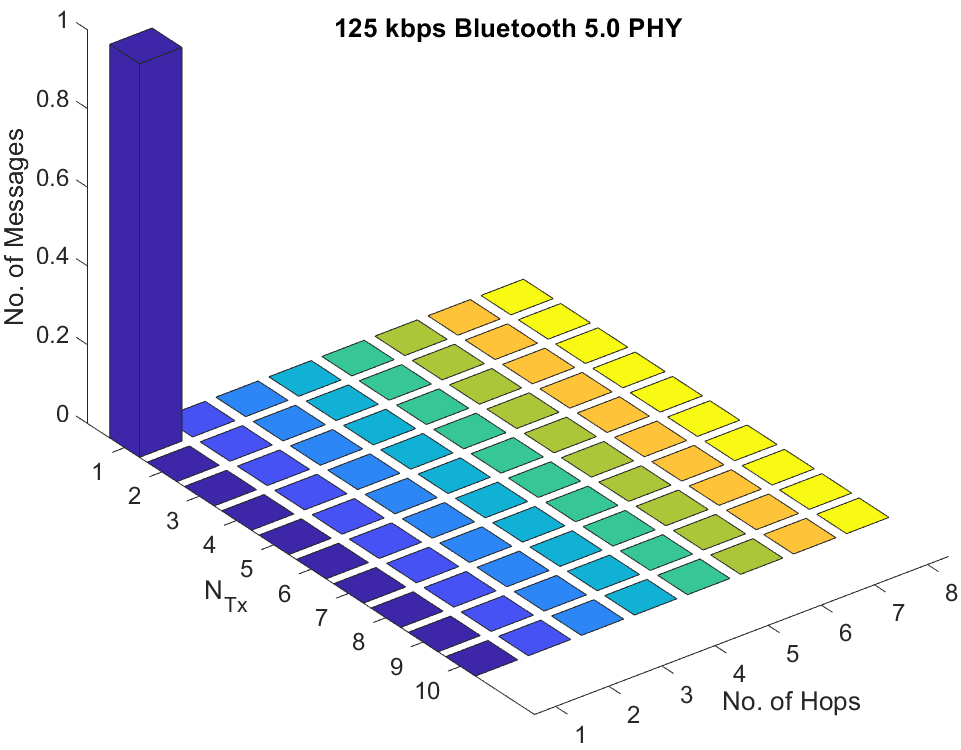}
    	\caption{}
    	\label{fig:sim_latency_125K}
    \end{subfigure}
    \begin{subfigure}[t]{0.32\textwidth}
    	\centering
    	\includegraphics[scale=0.2]{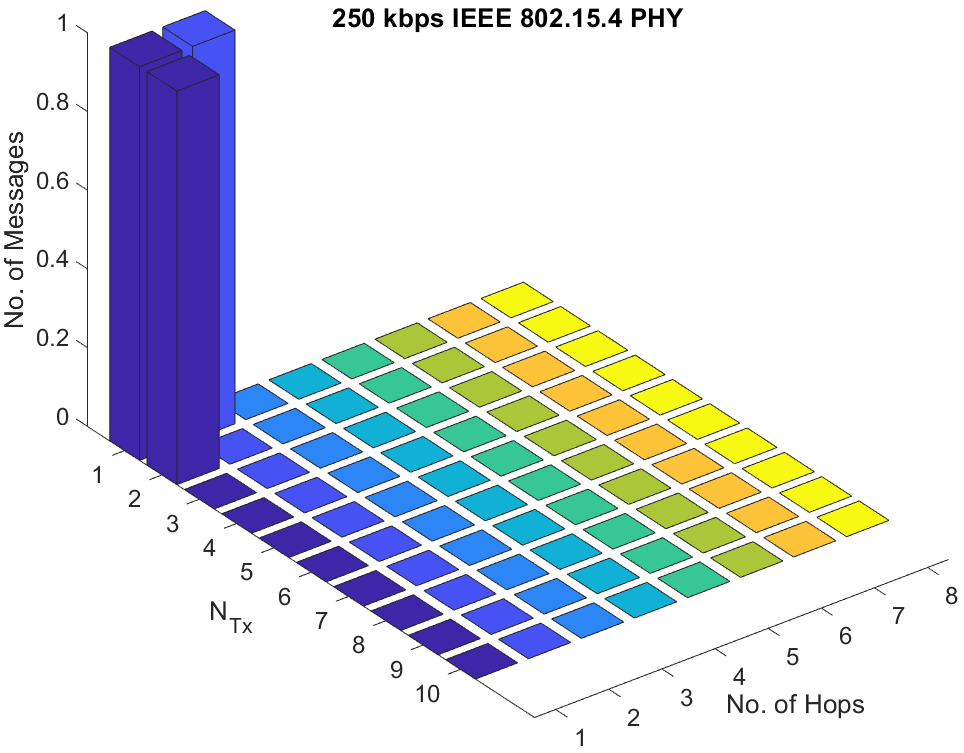}
    	\caption{}
    	\label{fig:sim_latency_154}
    \end{subfigure}
    \caption{Number of distinct messages that can be transmitted within a slotframe of 10 ms: (a) LE 2M PHY; (b) LE 1M PHY; (c) LE Coded PHY (500K); (d) LE Coded PHY (125K); (e) 802.15.4 PHY. }
    \label{fig:sim_latency}
\end{figure*}

\section{Experimental Validation}
\label{sec:experimentation}
Experiments were run with the \nrf platform on the \dcubee~\cite{schuss17competition} testbed. For the purposes of this paper, \sixpp uses a modified version of the \atomic CT scheduling architecture~\cite{baddeley2019atomic}, which has recently been extended to support CT-based protocols over the four \btfive PHY layers~\cite{baddeley2020impact} and incorporates recent \nrf support from Contiki-NG\footnote{https://github.com/contiki-ng/contiki-ng/pull/1310}. However, the \sixpp approach could also be replicated on other CT scheduling architectures such as Baloo~\cite{jacob2019baloo}. 

\boldpar{Association and destination advertisement}
Fig.~\ref{fig:eb_delta} shows the time taken for nodes to to receive their first EB (i.e., the latency of network association). In \sixpp, 6TiSCH's EB beaconing mechanism is replaced with centrally-broadcast EBs flooded over CT from the timesync/coordinator. As CT-based protocols intentionally contend and allow a message to be fully broadcast across the network with minimal latency, \sixpp nodes are able to synchronize and associate with the network in a fraction of the time required in traditional 6TiSCH networks, and furthermore allows nodes to directly synchronize with a single timesource (as opposed to their neighbor). 
Fig.~\ref{fig:dao_delta} shows delta between a node sending a DAO to receiving a DAO-ACK from the coordinator. In this case, while DAO's are still sent \emph{upwards} over the layer-3 RPL links, \sixpp again returns centrally computed DAO-ACKs as CT floods from the coordinator. By broadcasting DAO-ACKs in this fashion (as also demonstrated in~\cite{istomin2019route}), \sixpp reliably establishes RPL \emph{downward} routes more quickly than a hop-by-hop approach.

\boldpar{Reduction of control signaling}
\sixpp benefits from the network-wide reduction in control signaling. As EBs are sent over a CT flood, this becomes the default mechanism for synchronization, allowing \sixpp to completely dispense with 6TiSCH's KA messaging and thus freeing up the 6TiSCH slotframe. In Table~\ref{table:dcube_1hop} we examine the performance of \sixpp in \dcubee's dense data dissemination scenario. Furthermore, we employ JamLab-NG~\cite{schuss19jamlabng} to provide external narrowband radio interference. In this scenario, 64B messages are periodically generated at 5s intervals by the testbed and are disseminated from the coordinator to 20 nodes in a dense cluster. Even under the external interference conditions, both \sixpp and 6TiSCH demonstrate high reliability due to channel hoping mechanisms over both the CT flood and the 6TiSCH slotframe (as previously shown in Fig.~\ref{fig:6pp_example}). However, Fig.~\ref{fig:dcube_1hop} shows how the reduced control signaling in \sixpp (in the form of eliminating EB and KA beaconing) results in application-level messages experiencing fewer collisions within the slotframe. This improves \sixpp's end-to-end performance, resulting in a 42\% reduction in latency without interference, and 37\% under the external interference scenario.

\section{Simulation-based Evaluation}
\label{sec:simulation}

%\boldpar{Latency analysis}

%\todo{Adnan}
To complement our experimental investigation, we have conducted simulation-based evaluations of \sixpp's latency. The multi-PHY capabilities of \sixpp enable it to transmit multiple signaling/data messages within a single 6TiSCH slotframe. 

We assume that a CT-based flood carries a single signaling/data message from the controller to the entire multi-hop network.  The number of distinct messages at a given PHY within a 6TiSCH slotframe can be calculated as:
\begin{equation}
    \label{eq_floods}
    N_{Messages}^{PHY}=\left\lfloor \frac{T_{SF}}{T_{slot}^{PHY}\times \left(N_{Tx}+N_H \right)} \right\rfloor,
\end{equation}
where \(T_{SF}\) is the slotframe duration, \(T_{slot}^{PHY}\) is the slot duration for a CT, \(N_{Tx}\) is the number of times a message is transmitted after reception, and \(N_H\) is the (required) number of hops. 

Fig. \ref{fig:sim_latency} shows the number of distinct 64 bytes messages (as per the \dcubee experimentation) that can be transmitted within a standard 6TiSCH slot duration of 10ms. The evaluation is based on a radio ramp-up of 40 \(\mu\)s~\cite{nrf52840_productsheet} and a CT protocol overhead of 6 bytes. The results indicate that the fast data transmission capabilities of the BT\,5's uncoded physical layers provide an opportunity to reliably disseminate multiple messages over multiple hops when using CT-based flooding (in comparison to standard 6TiSCH) while the multi-message dissemination capability is somewhat limited at coded BT\,5 and IEEE 802.15.4 physical layers. 

% \boldpar{Reliability analysis}

% \todo{Adnan}
\section{Discussion and Future Work}
\label{sec:conclusions}
% This paper combines \btfive-based CT floods within 6TiSCH, a standard primarily based on the \ieee link-layer. 

This paper has validated the novel approach taken by \sixpp with initial promising results, and demonstrated the potential of \btfive-based CT flooding to support the IETF 6TiSCH standard. The synchronicity of CT flooding allows messages to be sent in a deterministic manner that fits neatly into the 6TiSCH slotframe, and Sect.~\ref{sec:experimentation} showed how carrying EBs and DAO-ACKs over CT-floods provides a low-latency mechanism for centralized control signaling: freeing the slotframe for layer-3 messaging. 

Indeed, existing mechanisms in the primarily \ieee-based 6TiSCH standard seem to be amenable to the inclusion of CT-based floods. For example, \emph{hard cells} can be used to reserve immutable slots for CT floods. Unlike \emph{soft cells}, these cannot be altered on an ad-hoc basis, which means that the operation of CT floods can be neatly decoupled from 6TiSCH. This would help to support the SDN~\cite{ietf_sdn} concepts that are included as part of the standard. 

As demonstrated in Sect.~\ref{sec:simulation}, \btfive-based CT flooding (as opposed to \ieee CT flooding) can disseminate data at a far greater rate than currently possible within \ieee-based 6TiSCH. While \ieee-based flooding would be a more universal approach, we believe adopting the multi-PHY capabilities of modern low-power wireless chipsets brings significant gains, and opens the possibility of adaptively switching the CT PHY at runtime in order to adapt to changing network conditions. Indeed, such an approach has been proposed in recent works that have expanded our understand of CT communications~\cite{alnahas2019concurrentBLE5,baddeley2020impact} and is gaining considerable attention within the community. When utilizing the \btfive high data-rate uncoded PHYs \sixpp could also provide a means for rapidly distributing larger IPv6 packets such as those required in firmware updates -- significantly reducing the time taken to update a network. In short, the gains that can be achieved with CT flooding protocols over the \btfive PHY layers provides a strong argument for focusing future standardization efforts on interoperability between the two (CT and 6TiSCH). 

Finally, \sixpp opens other intriguing directions for future research -- particularly in mobility and synchronization. Firstly, by eschewing the need for topology control, CT protocols naturally lend themselves towards mobility scenarios. The addition of CT-based control mechanisms within the 6TiSCH slotframe could therefore support more dynamic scenarios than the current industrial IoT use-cases envisioned by the standard. Additionally, 6TiSCH currently requires generous guards to tolerate drifting~\cite{elsts2016microsecond}. This drift tolerance subsequently dictates the length of a TSCH timeslot, not only representing a minimum bound on latency but also incurring a cost on energy. As \sixpp maintains highly accurate time synchronization through the CT scheduler, there is scope for reducing these guard times to provide even lower end-to-end latencies.

\vspace{+1mm}
% \todo{Switching the CT PHY based on application/environment.}

% use section* for acknowledgements
% \section*{Acknowledgements}

% Acknowledgments go here

% References section
 \bibliographystyle{IEEEtran}

\iffalse
\bibliography{references,baddeley}
\else
\bibliography{references,blind}
\fi

\end{document}